\documentclass{article}
\usepackage[accepted]{vietnam} 
\usepackage{natbib}
\usepackage{graphicx}
\usepackage{subfigure}      
\newcommand{\unit}[1]{\ensuremath{\, \mathrm{#1}}}
\newcommand{\ramses}{{\sc ramses}}          

\newcommand{\Fig}[1]{Fig.\ \ref{fig:#1}}    

\begin{document}
\twocolumn[
\title{Accounting for the diversity in stellar environments}
\titlerunning{Accounting for the diversity in stellar environments}
\author{Michael K\"uffmeier}{kueffmeier@nbi.ku.dk}
\address{Centre for Star and Planet Formation, Niels Bohr Institute and Natural History Museum of Denmark, University of Copenhagen,
{\O}ster Voldgade 5-7, DK-1350 Copenhagen K, Denmark}
\author{Troels Haugb{\o}lle}{haugboel@nbi.ku.dk}
\address{Centre for Star and Planet Formation, Niels Bohr Institute and Natural History Museum of Denmark, University of Copenhagen,
{\O}ster Voldgade 5-7, DK-1350 Copenhagen K, Denmark}
\author{{\AA}ke Nordlund}{aake@nbi.ku.dk}
\address{Centre for Star and Planet Formation, Niels Bohr Institute and Natural History Museum of Denmark, University of Copenhagen,
{\O}ster Voldgade 5-7, DK-1350 Copenhagen K, Denmark}

\vskip 0.5cm 
]

\begin{abstract}
Stars and their corresponding protoplanetary disks form in diverse environments.
To account for these natural variations, 
we investigate the formation process around nine solar mass stars with a maximum resolution of
2 AU in a Giant Molecular Cloud of (40 pc)$^3$ in volume by using the adaptive mesh refinement
code \ramses. 
The magnetohydrodynamic simulations reveal that the accretion process is heterogeneous in time, 
in space, and among protostars of otherwise similar mass. 
During the first roughly 100 kyr of a protostar evolving to about a solar mass,
the accretion rates peak around $10^{-5}$ to $10^{-4}$ M$_{\odot}$ yr$^{-1}$ 
shortly after its birth, declining with time after that.
The different environments also affect the spatial accretion, and infall of material to the 
star-disk system is mostly through filaments and sheets. 
Furthermore, the formation and evolution of disks varies significantly from star to star. 
We interpret the variety in disk formation as a consequence of the differences in 
the combined effects of magnetic fields and turbulence 
that may cause differences in the efficiency of magnetic braking,
as well as differences in the strength and distribution of specific angular momentum.
\end{abstract}

\section{Introduction}
Protoplanetary disks form around stars as a consequence of pre-stellar cores
collapsing in filaments of Giant Molecular Clouds,
which makes them the smallest entity in a hierarchy of scales.
Length scales range from tens of parsecs for Giant Molecular Clouds to
protoplanetary disk sizes of $\sim 10$ AU to $\sim 100$ AU. It is computationally
very challenging to cover such a broad range of scales in a single simulation.
Therefore, simulations of protostellar formation traditionally start from initial conditions
representing a collapsing spherically symmetric cloud, as an approximation to the pre-stellar core
\citep{2004MNRAS.348L...1M,2006ApJ...645.1227M,2007ApJ...670.1198M,2011MNRAS.413.2767M,2012A&A...543A.128J,2013A&A...554A..17J,2010ApJ...714L..58T,2013ApJ...763....6T,2011ApJ...738..180L,2011MNRAS.417.1054S,2012MNRAS.422..347S,2016arXiv161003324V}.
This approach allows detailed parameter studies, but
neglecting the underlying turbulence in Giant Molecular Clouds and the potential interactions with the surroundings
could potentially limit the applicability of such idealized initial conditions.
Considering the dynamics of Giant Molecular Clouds, it is important to
investigate how they affect the formation of protostars and protoplanetary disks. 
Given that most of the volume in the Giant Molecular Cloud
is of relatively low density and thus of less interest for star formation,
simulating the huge range of scales becomes feasible by applying adaptive mesh
refinement to the problem.
First, we briefly explain the concept behind our zoom-method, 
which allowed us to resolve the accretion and disk formation process, 
while simultaneously accounting for the large-scale environment.  
Second, we present an overview of the most significant results obtained in our study
before we discuss and summarize their consequences. 

\begin{figure}
\subfigure{\includegraphics[width=\linewidth]{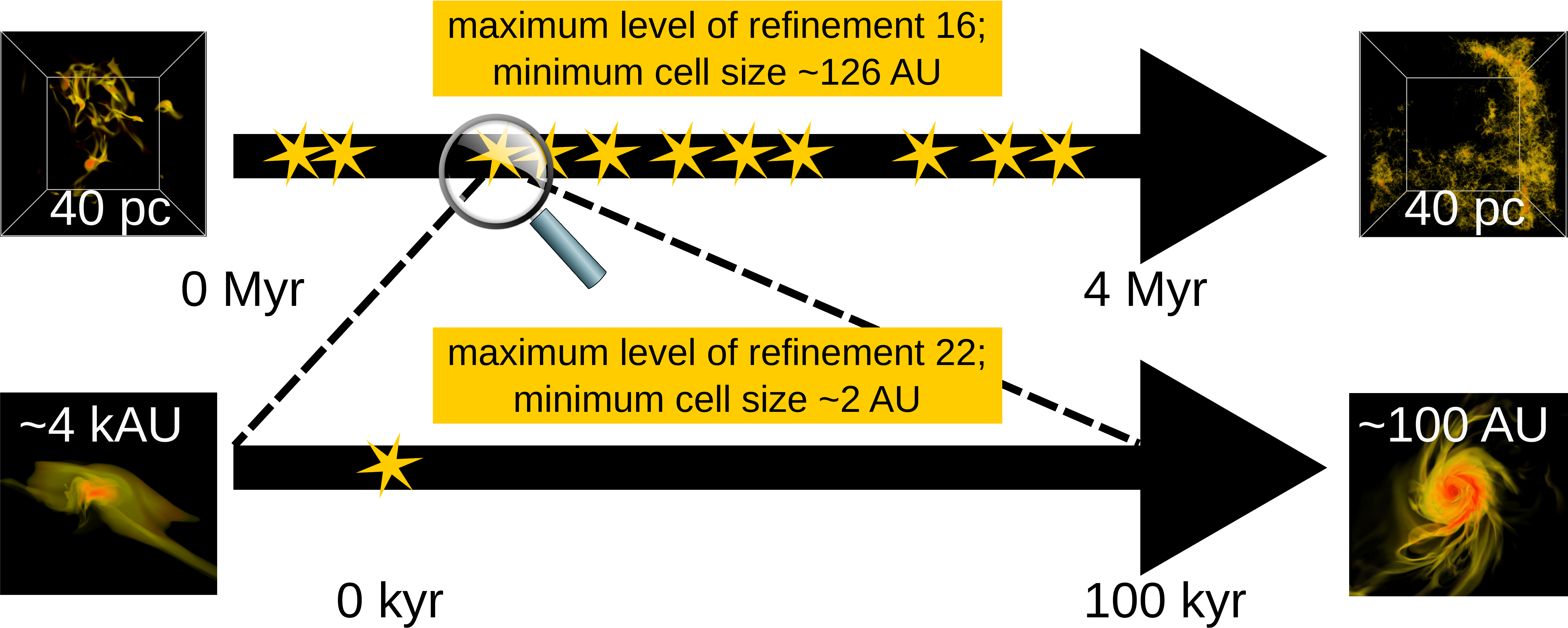}} 
\protect\caption{\label{fig:time-sketch} Sketch of the zoom-in procedure.
First we evolve a snapshot of an already turbulent
GMC of $(40 \unit{pc})^3$ in size (upper left image) for about 4 Myr (upper right image).
During the evolution multiple sinks are created and
at the current time more than 500 stars have formed.
We zoom in on selected pre-stellar cores (lower left image) to resolve the
formation process with higher resolution for up to about 100 kyr after sink creation (lower right image).
This procedure is applied to altogether nine protostars.}

\end{figure}

We use use a highly modified
version of the adaptive mesh-refinement code \ramses\ \citep{2002A&A...385..337T,2006A&A...457..371F},
which in principle can handle refinement over up to 29 factors of two \citep{2014IAUS..299..131N}.
In \Fig{time-sketch}, we sketch the procedure and refer the reader to 
\citep{2016ApJ...826...22K,kuffmeier17a,kuffmeier17b} for further details.

We start from an already turbulent GMC model of a cubic box of size (40 pc)$^3$ with 
periodic boundary conditions,
consisting of self-gravitating, magnetized gas. 
The average $H_{2}$ number density is $30$ cm$^{-3}$,
which yields a total mass of the box of approximately $10^{5}$M$_{\odot}$.
The assumed GMC lifetimes are in agreement with the 'star formation in
a crossing time' paradigm \citep{2000ApJ...530..277E,2003MNRAS.338..817E,2016ApJ...822...11P},
and with observational estimates \citep{2011ApJ...729..133M}, and the turbulence is driven
by massive stars that inject energy of
$10^{51}$ erg of thermal energy into the GMC after a mass-dependent life-time.
For the heating via UV-photons \citep{2006agna.book.....O}, 
we apply the recipe of \citep{1986PASP...98.1076F}
and use an optically thin cooling function
\citep{2012ApJS..202...13G} for the cold dense gas.

The combined effects of turbulence and self-gravity induce the formation of 
filaments, and subsequently star formation inside the filaments. 
To make the problem computationally tractable, we describe the collapse of matter into stars with a sub-grid sink particle algorithm. 
As illustrated in \Fig{time-sketch}, we first evolve the GMC with a
minimum cell size of 126 AU before we zoom-in onto the individual sinks of interest
with a minimum cell-size of 2 AU.
This second stage provides information
about protostellar accretion, including the
subsequent formation of protoplanetary disks.

\section{Results of the zoom-ins}
\begin{figure}
\subfigure{\includegraphics[width=\columnwidth]{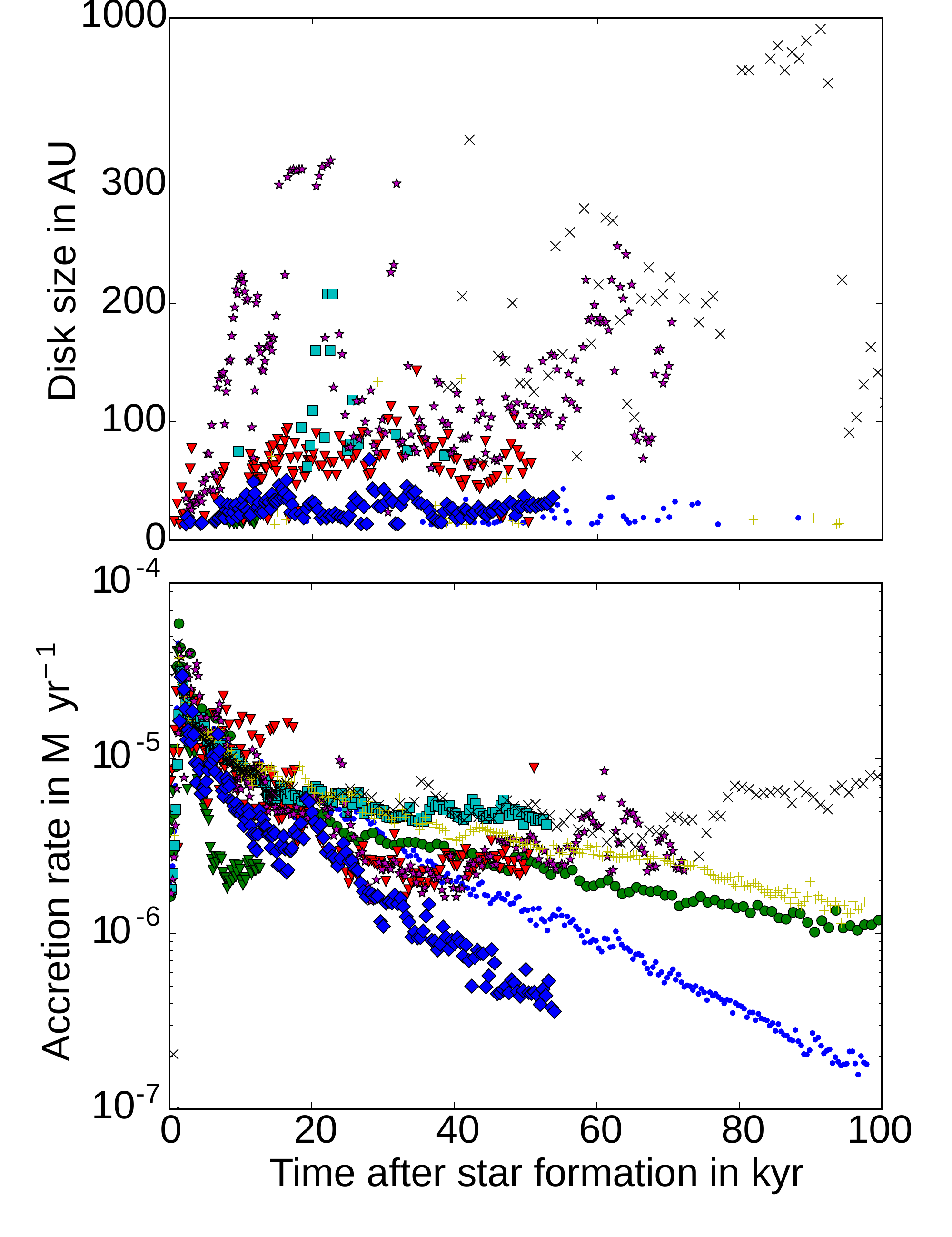} }
\protect\caption{\label{fig:disk_acc} Upper panel: Evolution of the disk radius
around the different sinks.
Lower panel: 
Accretion profile for 9 sinks created in zoom-ins started 
with increased resolution before sink creation. 
}
\end{figure}

The different environments of the protostars cause differences in the accretion process
and disk formation among the protostars. 
We illustrate the accretion profiles of nine sinks in the lower panel of \Fig{disk_acc}.
One can see a general trend for the different sinks, with a very steep initial increase to values of about
$10^{-4} \unit{M}_{\odot}\textrm{yr}^{-1}$ to $10^{-5} \unit{M}_{\odot}\textrm{yr}^{-1}$
followed by a general decrease. 
The decrease varies between sinks and some of the sinks still show accretion rates 
of more than $10^{-6} \unit{M}_{\odot}\textrm{yr}^{-1}$ after $\sim 100$ kyr. 
Moreover, we can see that some of the sinks show significant fluctuations during their evolution.
Since we average over 
periods of 200 to 400 years between the snapshots we are underestimating the amplitude of these
episodic accretion events. 
Finally, we note that the sinks accrete their mass through accretion channels \citep{2013MNRAS.432.3320S}
rather than uniformly in space.

\begin{figure}
\subfigure{\includegraphics[width=\columnwidth,viewport=100bp 0bp 602bp 580bp,clip=true]{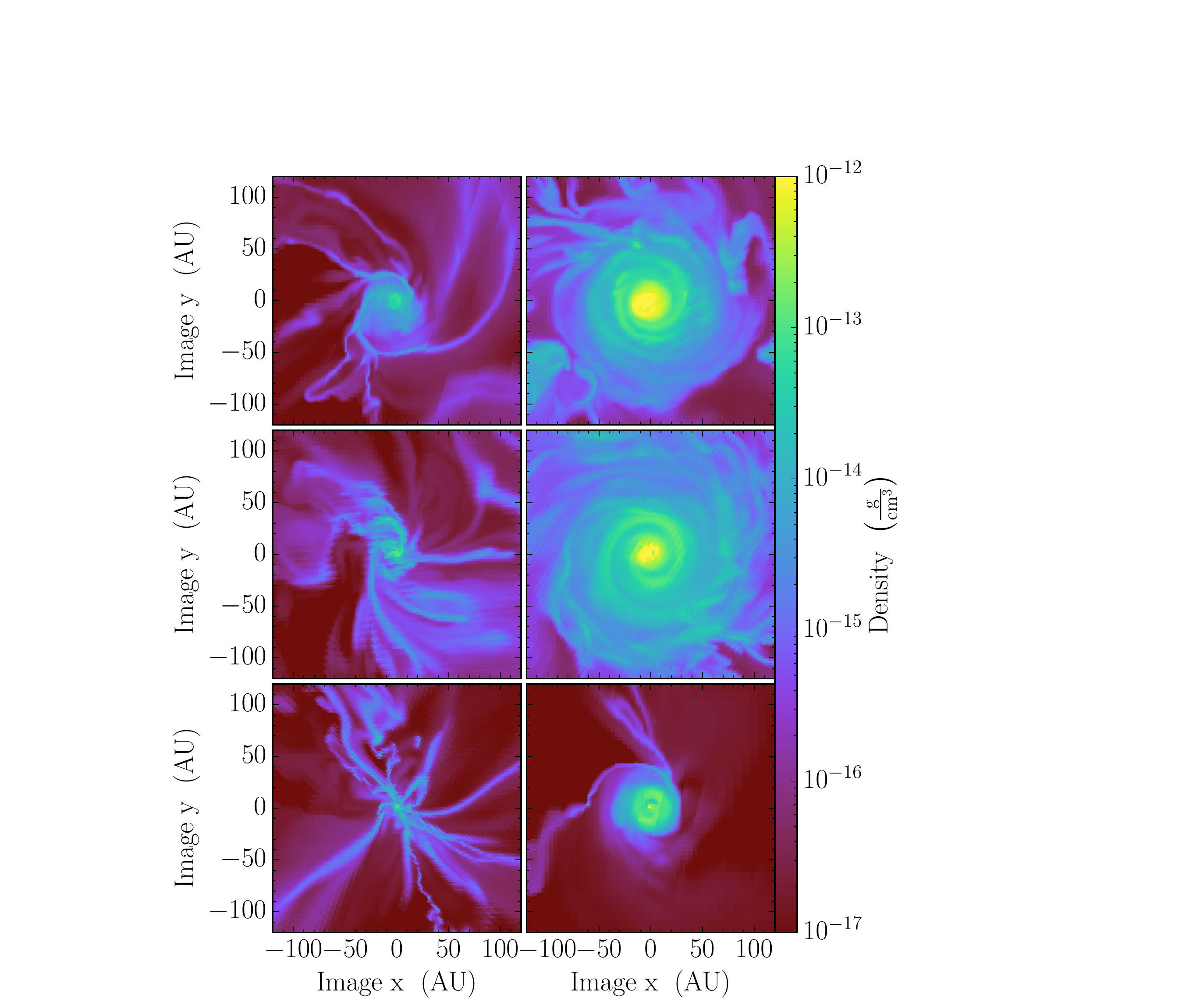} } 
\protect\caption{\label{fig:disk-slices} 
Slices in the plane vertical to the mean angular momentum vector
calculated for a sphere of 100 AU around six different sinks at $t=50$ kyr.
The upper left slice corresponds to the sink with the blue dots in the accretion plot,
the upper right to the red triangles, the middle left to the cyan squares,
the middle right to the magenta asterisks, the lower left to the yellow pluses,
and the lower right to the blue diamonds.
}
\end{figure}

\Fig{disk-slices} shows slices in the plane 
perpendicular to the mean angular momentum vector
at $t=50$ kyr around six sinks.
The images and the upper panel in \Fig{disk_acc} reveal the variety in disk formation for the different stellar environments,
and also the spatial variations in the accretion process induced by filamentary arms feeding the forming protoplanetary disk.
Also, the disks show signs of spiral arms or inflowing gas streams 
strikingly similar to what has been observed by ALMA
and with the Subaru Next Generation Adaptive Optics (HiCIAO) \citep{Liue1500875}.

\section{Discussion and Conclusion}

Using a numerical model that simultaneously encompass the large-scale environment of a Giant Molecular Cloud 
and the the immediate environment of nine protostars, covering seven orders of magnitude in dynamic range, 
we have investigated the environmental effects on the protostellar formation process.
One major result is that stellar accretion can be very different depending on the protostellar environment.
We also conclude that the diversity in the large-scale stellar environment profoundly
influences the formation and evolution of protoplanetary disks.

If the magnetization of the surrounding gas is sufficiently limited to avoid the magnetic braking catastrophe,
protoplanetary disks of several tens of AU can form as early as a few thousand years after star formation.
In cases where the magnetization of the collapsing gas is sufficiently large (low mass-to-flux ratios),
no disk of more than $\approx$10 AU in size will form around the star.
The main reason why the magnetic braking catastrophe is avoided in many cases is the reduction of magnetic
braking caused by turbulence.

\section*{Acknowledgments}
This research was supported by a grant from the Danish Council for Independent Research to {\AA}N,
a Sapere Aude Starting Grant from the Danish Council for Independent Research to TH.
Research at Centre for Star and Planet Formation is funded by the Danish National Research Foundation (DNRF97).
We acknowledge PRACE for awarding us access to the computing resource CURIE based in France at CEA for
carrying out part of the simulations.
Archival storage and computing nodes at the University of Copenhagen HPC center, funded with a research
grant (VKR023406) from Villum Fonden, were used for carrying out part of the simulations and the post-processing. 
Finally, we acknowledge the developers of the 
python-based analyzing tool yt (http://yt-project.org/) \citep{2011ApJS..192....9T} that simplified our analysis.

\bibliography{SFDEbibexample}

\end{document}